\documentclass[12pt, draftclsnofoot, onecolumn]{IEEEtran}
\ifCLASSINFOpdf
\else
\fi
\usepackage{color}
\usepackage{cite}
\usepackage[cmex10]{amsmath}
\usepackage{amsthm}
\usepackage{algorithm}
\usepackage{algorithmic}
\usepackage{stfloats}
\usepackage{subfigure}
\usepackage{subfig}
\usepackage{mathrsfs}
\usepackage{multicol,multienum}
\usepackage[pdftex]{graphicx}

\begin{document}
%

\title{Network-Connected UAV Communications: Potentials and Challenges}

\author{Haichao Wang, Jinlong Wang, Jin Chen, Yuping Gong, and Guoru Ding

%
\thanks{This work has been accepted by China Communications.}
\thanks{This work is supported by the National Natural Science Foundation of China (Grant No. 61501510), Natural Science
Foundation of Jiangsu Province (Grant No. BK20150717), China Postdoctoral Science Funded Project, and Jiangsu Planned Projects for Postdoctoral Research Funds.} 
\thanks{The authors are with College of Communications Engineering, Army Engineering University of PLA, Nanjing 210007, China. G. Ding is also with National Mobile Communications Research Laboratory, Southeast University, Nanjing 210096, China (e-mail: dr.guoru.ding@ieee.org).}}

\IEEEpeerreviewmaketitle
\maketitle
\begin{abstract}
This article explores the use of network-connected unmanned aerial vehicle (UAV) communications as a compelling solution to achieve high-rate information transmission and support ultra-reliable UAV remote command and control. We first discuss the use cases of UAVs and the resulting communication requirements, accompanied with a flexible architecture for network-connected UAV communications. Then, the signal transmission and interference characteristics are theoretically analyzed, and subsequently we highlight the design and optimization considerations, including antenna design, nonorthogonal multiple access communications, as well as network selection and association optimization. Finally, case studies are provided to show the feasibility of network-connected UAV communications.
\end{abstract}


\IEEEpeerreviewmaketitle
\section{Introduction}
The dramatically growing demand for high-rate and ubiquitous wireless communication services has impelled the unmanned aerial vehicle (UAV) communications to be an active research area recently\cite{CMZY,CMDGR,CMWHC,TWCMM,TWCMM2}. Benefiting from the high maneuverability, the UAV can be quickly deployed to provide wireless services for some hotspots and in case of terrestrial base station (BS) failure. Besides, the flexible location provides additional performance gains compared with fixed infrastructure based communications\cite{WCLWHC,TWCWQQ}. However, these advantages also suffer from many challenges. In particular, existing UAV communication systems are mainly based on the direct ground-to-UAV communications over the unlicensed spectrum or reuse the spectrum bands that have been assigned for other particular applications\cite{CMZY}, which results in limited data rate, unreliable connections, and insecure communications. In this context, one challenging task is to establish high-rate and reliable communication links with UAVs.

Network-connected UAV communications, where UAVs are connected to the terrestrial cellular and satellite networks, have received increasing research attention since today's cellular and satellite networks are almost ubiquitous accessibility worldwide\cite{AccessXZ,GloMMA,ICCMMA,WCLRA,CellularZY,CellularZY2}. Authors in\cite{AccessXZ} investigate a joint time-frequency scheduling and power optimization, where multiple UAVs are controlled by a terrestrial BS. In\cite{CellularZY}, to maintain reliable wireless connection with the cellular network by associating with one of the ground BSs, the UAV's trajectory optimization is studied. Additionally, the feasibility of using the existing cellular infrastructure for supporting UAV communications is analyzed in\cite{GloMMA,ICCMMA}. Other studies include radio channel modeling for UAV communications over cellular networks\cite{WCLRA}, UAV-aided cellular offloading\cite{CellularZY2}, to name just a few.

The network-connected UAV communications, which are expected to achieve high-rate information transmission and ultra-reliable UAV remote control, are of great importance but largely unexplored. This article aims to elaborate the design aspects and open issues in network-connected UAV communications. In particular, we first discuss the use cases of UAVs and the resulting communication requirements. Then, we propose a flexible architecture for network-connected UAV communications. Further, the signal transmission and interference characteristics are theoretically analyzed, and subsequently we highlight the design and optimization considerations, including antenna design, nonorthogonal multiple access (NOMA) communications, as well as network selection and association optimization. Moreover, case studies are provided to show the feasibility of network-connected UAV communications.
\section{Requirements and Architecture}
\subsection{Use Cases and Requirements}
\begin{figure}[!t]
\centering{\includegraphics[width=100mm]{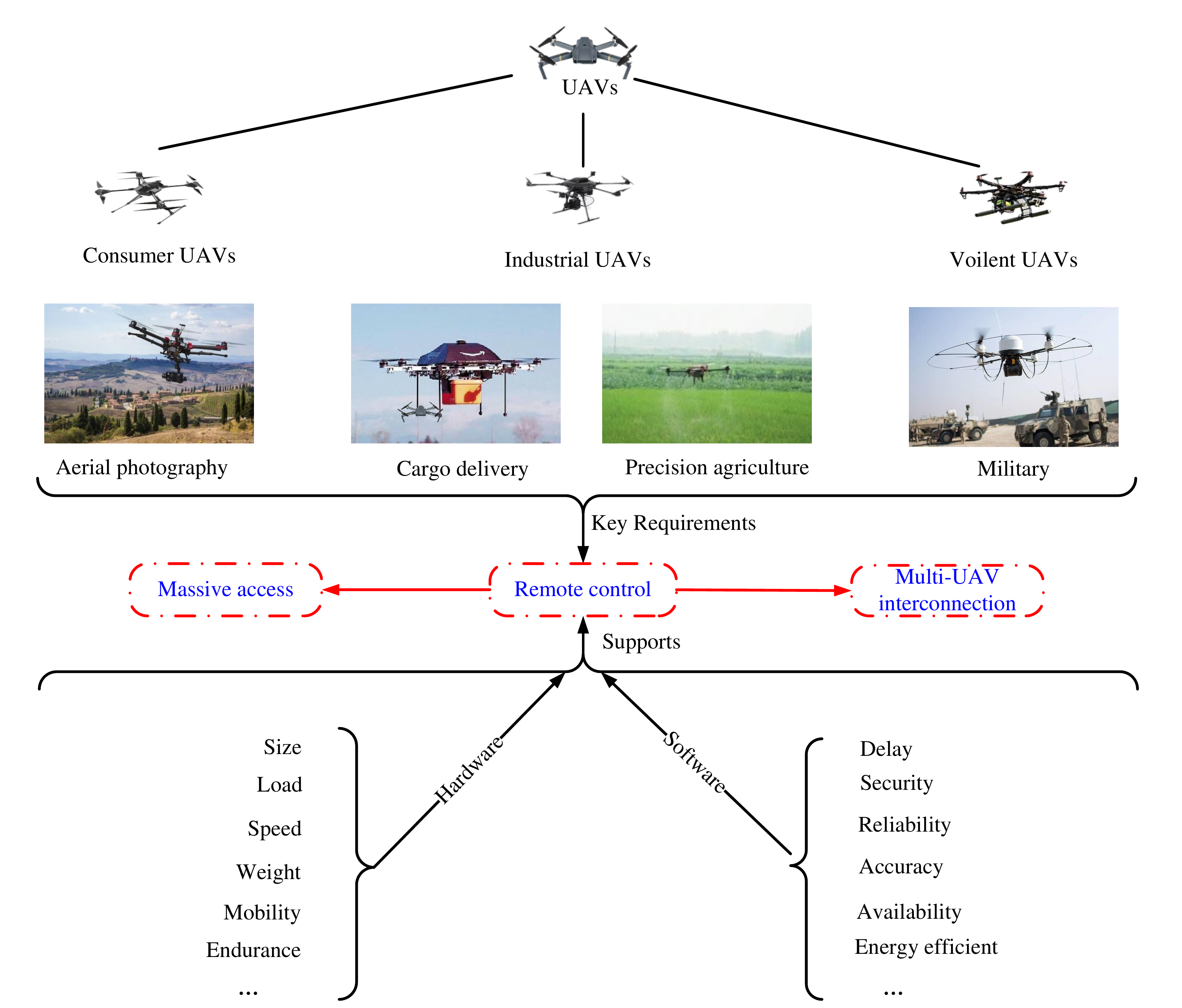}}
\caption{The key requirements for supporting various UAV use cases.}
\label{Applications}
\end{figure}

UAVs, also known as ``drones'', vary significantly in size from small toys to large military aircrafts\cite{UAVSize}. In this article, the focus is on the small UAVs. As shown in Fig. \ref{Applications}, according to their use cases, UAVs can be categorized into three groups: i) consumer UAVs, where UAVs are used mainly by individuals. A representative example is for aerial photography; ii) industrial UAVs, where UAVs are employed by enterprises or organizations to perform specific tasks, such as cargo delivery, disaster relief, precision agriculture, engineering inspections, and communication relaying\cite{TCZY}, to name a few; iii) voilent UAVs, where UAVs are designed for police operations and military affairs.
The myriad of possible scenarios in which the UAV plays an important role necessitates the development of UAV communications. In order to enable massive UAVs to orderly flight in the sky, the following key requirements must be satisfied.
\begin{itemize}
  \item \emph{Remote control}, where UAVs can be commanded and controlled by remote users without the limitions of line-of-sight (LOS) operation range. In practice, it is not guaranteed that the UAV is always within the LOS range when performing any missions, such as cargo delivery. Achieving remote command and control is the key to efficiently fulfill various tasks.
  \item \emph{Massive access,} where massive UAVs are expected to access the network and be controlled simultaneously. It is predicted that more than 7 million consumer UAVs will operate across Europe in 2050\cite{UAVnumber}. Therefore, massive access for UAVs must be supported, which is the foundation of UAV systems' running.
  \item \emph{Multi-UAV interconnection}, where multiple UAVs can exchange information by directly point-to-point communications or information relaying, so as to inform other UAVs about their or network current status. When multiple UAVs fly in the sky, they must plan their trajectories in real time to avoid collisions. In this process, one of extremely important issues is information exchange. Multi-UAV interconnection is the guarantee of realizing information exchange.
\end{itemize}
To support these stringent requirements, specific metrics are expected to be achieved, as illustrated in Fig. \ref{Applications}. From the software perspective, high data rate, low delay, high security, etc., should be considered. Meanwhile, the hardware techniques are required to be greatly developed, such as high load, small size, and long endurance.
\subsection{Networking Architecture}
\begin{figure}[!t]
\centering{\includegraphics[width=130mm]{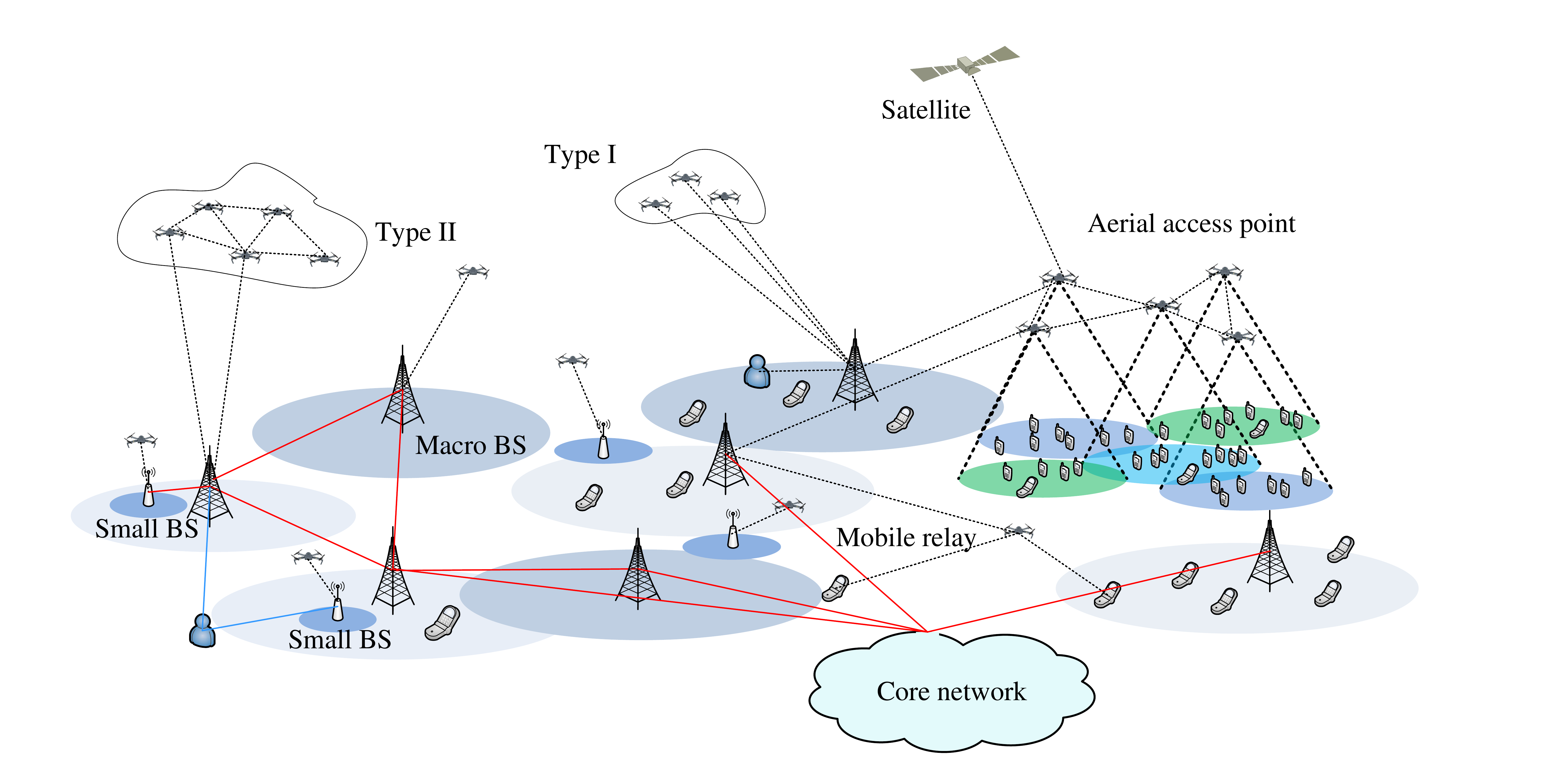}}
\caption{The networking architecture for network-connected UAV communications.}
\label{Architecture}
\end{figure}
Satisfying the aforementioned requirements is challenging due to the highly dynamic topology, the high speed of the UAV, and heterogeneous quality of service (QoS) requirements (e.g., the asymmetric QoS requirements for downlink and uplink communications). An adequate choice is integrating UAVs into the cellular (or long term evolution) and satellite networks to establish a reliable wireless connectivity for UAV applications, also termed as network-connected UAV communications. In this paper, our focus is on the cellular network-connected UAV communications. To achieve remote control, massive access and multi-UAV interconnection, the proposed networking architecture is shown in Fig. \ref{Architecture}. First of all, there are two forms of communications for UAVs to access the network.   One is that each UAV directly communicates with the BS, i.e., type I in Fig. \ref{Architecture}. The other is that the UAV is connected to the network through the aerial relays, i.e., type II in Fig. \ref{Architecture}, where some UAVs act as gateways that perform direct UAV-to-ground (ground-to-UAV) communications. Moreover, the networking architecture shows significant heterogeneity since the access points can be either macrocell BSs or smallcell BSs. The types of UAVs are also diverse.

As can be seen from Fig. \ref{Architecture}, the key requirements can be effectively achieved by implementing network-connected UAV communications. Firstly, there are a large number of BSs, especially in which ultra dense networks are deployed. These BSs can provide network connections for massive UAVs while serving ground users. Therefore, massive access for UAVs can be supported. Moreover, ubiquitous BSs can be interconnected through the core network or other interfaces. This means that via accessing the network, a user can command and control a network-connected UAV that is far away from the user. This is no longer the traditional direct ground (or UAV)-to-UAV communications. It's the ground (or UAV)-to-network-to-UAV communications. Additionally, multiple UAVs can also be interconnected effectively by two forms of communications: UAV-to-UAV and UAV-to-network-to-UAV. The best strategy in choosing different models depends on many factors, such as the locations of UAVs, available spectrum, on-board energy, and communication requirements, etc. For example, if there is spectrum available and the communication link between two UAVs is of good quality, UAVs can communicate directly without the assistance of network. When the UAV's transmit power is low due to lack of energy, the UAV can select nearby UAVs or terrestrial networks to forward the information.

The terrestrial communications in return can be empowered by employing UAVs. As an aerial access point, UAV can provide network access for ground users. As a mobile relay, the UAV can forward information among users without reliable direct links. Compared with traditional communications based on fixed infrastructure, UAV-assisted communications can achieve additional performance gains by dynamically adjusting its locations\cite{CMZY,CMWHC,TWCWQQ}. At present, the UAVs mainly can be classified two groups: tethered and untethered UAVs. A tethered UAV is connected by a cable/wire with the ground control platform, thus it has stable power supply. In this case, UAVs can work without interruption. On the other hand, the untethered UAV must rely on its on-board energy. The UAV must return for charging when insufficient energy is warned. Since there are multiple UAVs, the system can successfully run if one of UAVs runs out of power because it can be substituted by others. The development of UAV communications relies heavily on the advances in hardware technology. Because of the hardware limitations, UAVs may not effectively exert their performance in some harsh environments, such as strong winds and hail. Therefore, the terrestrial communication is an indispensable means.
\section{Signal Transmission and Interference Characteristics}
The communication links in network-connected UAV communications consist of three kinds of channels: Ground-to-UAV, UAV-to-ground and UAV-to-UAV channels. All these channels show several characteristics compared with the terrestrial communication channels. The UAV-to-ground and UAV-to-UAV channels have been discussed and studied in\cite{CMZY,UAVSize}. The focus of this article is on the ground-to-UAV channel and associated signal transmission characteristics, based on which we will analyze the interference characteristics.
\subsection{Signal Transmission Characteristics}
For an aerial UAV, the received signal power is ${p_r} = {p_t}{g_t}{g_r}{g_c}$ with the transmit power ${p_t}$, transmit antenna gain ${g_t}$, receive antenna gain ${g_r}$, and channel power gain ${g_c}$ between the ground BS and the UAV. The ground-to-UAV channel ${g_c}$ can be modeled by the LOS and non-line-of-sight (NLOS) links separately along with their probabilities of occurrence\cite{GloMMA,ICCMMA}, i.e., ${g_c} = {\text{P}_L}{G_L}{d^{ - {\alpha _L}}} + (1 - {\text{P}_L}){G_N}{d^{ - {\alpha _N}}}$, where ${\text{P}_L}$ is probability of having LOS link, ${G_L}$ and ${G_N}$ are constants representing the path losses at the reference distance $d_0$, ${{\alpha _L}}$ and ${{\alpha _N}}$ are the path loss exponents for the LOS and NLOS links, and $d$ is the distance between the BS to the UAV. The probability of having LOS link between a BS with height ${h_{BS}}$ and a UAV with height ${h_{UAV}}$ is given by\cite{GloMMA}
\begin{align}
\label{Probability}
{\text{P}_L} = \prod\limits_{n = 0}^m {\left[ {1 - \exp \left( { - \frac{{{{\left[ {{h_{BS}} - \frac{{\left( {n + 0.5} \right)\left( {{h_{BS}} - {h_{UAV}}} \right)}}{{m + 1}}} \right]}^2}}}{{2{c^2}}}} \right)} \right]},
\end{align}
where $m = \left\lfloor {\frac{{r\sqrt {ab} }}{{1000}} - 1} \right\rfloor $, $r$ is the horizontal distance between the BS and the UAV, $a$, $b$ and $c$ are parameters that characterize the environment. It can be seen from (\ref{Probability}) that the probability not only relates to the environment, but also depends on the heights and horizontal distance of the UAV and BS. Increasing the UAV's height may acquire higher LOS probability. The LOS probability approaches 1 with sufficiently high altitude. On the other hand, this inevitably results in more serious path loss since the distance becomes larger.
 \begin{figure}[!t]
\centering{\includegraphics[width=120mm]{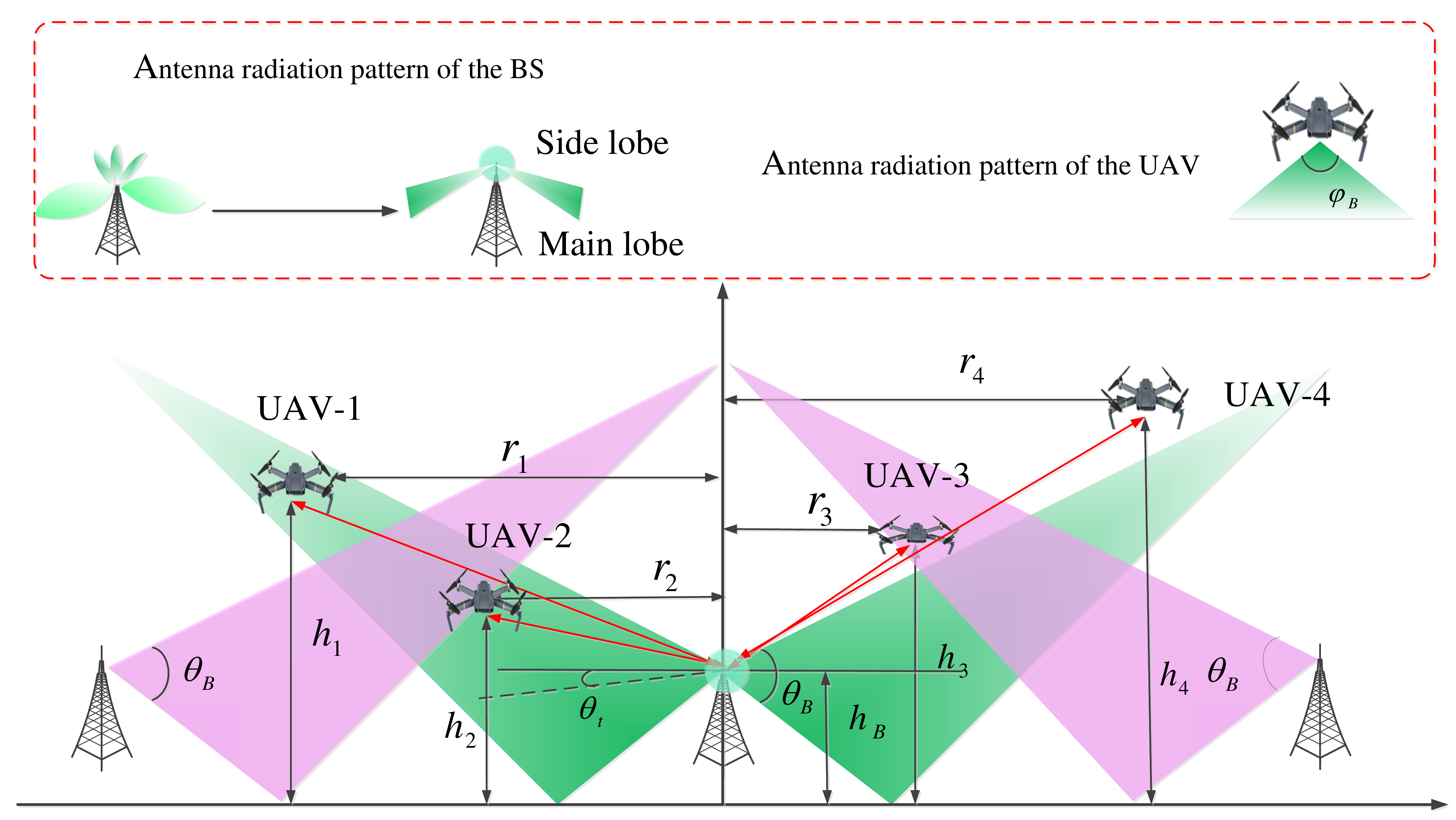}}
\caption{Illustrations of the BS's antenna and interference cases.}
\label{Channel}
\end{figure}

The received power also relies on the antenna radiation patterns of BSs and the UAV, which results in different transmit antenna gains $g_t$ and receive antenna gains $g_r$. Without loss of generality, consider that the antenna of the UAV is directional pointing directly downwards, and hence the antenna has a beamwidth of ${\varphi _B}$, as shown in Fig. \ref{Channel}. It can be seen that the antenna is undirectional in the case of ${\varphi _B} = 180^o$. Given the beamwidth, the receive antenna gain is about ${g_r} = {{29000} \mathord{\left/
 {\vphantom {{29000} {{\varphi _B}^2}}} \right.
 \kern-\nulldelimiterspace} {{\varphi _B}^2}}$. Moreover, to receive the information signal from the BS, the horizontal distance $r$ must satisfy $r \le \left( {{h_{UAV}} - {h_{BS}}} \right)\tan \left( {{{{\varphi _B}} \mathord{\left/
 {\vphantom {{{\varphi _B}} 2}} \right.
 \kern-\nulldelimiterspace} 2}} \right)$ or written as ${r \mathord{\left/
 {\vphantom {r {\left( {{h_{UAV}} - {h_{BS}}} \right) \le \tan \left( {{{{\varphi _B}} \mathord{\left/
 {\vphantom {{{\varphi _B}} 2}} \right.
 \kern-\nulldelimiterspace} 2}} \right)}}} \right.
 \kern-\nulldelimiterspace} {\left( {{h_{UAV}} - {h_{BS}}} \right) \le \tan \left( {{{{\varphi _B}} \mathord{\left/
 {\vphantom {{{\varphi _B}} 2}} \right.
 \kern-\nulldelimiterspace} 2}} \right)}}$ if ${h_{UAV}} > {h_{BS}}$ and ${r \mathord{\left/
 {\vphantom {r {\left( {{h_{UAV}} - {h_{BS}}} \right) \le \tan \left( {{{{\varphi _B}} \mathord{\left/
 {\vphantom {{{\varphi _B}} 2}} \right.
 \kern-\nulldelimiterspace} 2}} \right)}}} \right.
 \kern-\nulldelimiterspace} {\left( {{h_{UAV}} - {h_{BS}}} \right) > \tan \left( {{{{\varphi _B}} \mathord{\left/
 {\vphantom {{{\varphi _B}} 2}} \right.
 \kern-\nulldelimiterspace} 2}} \right)}}$ if ${h_{UAV}} < {h_{BS}}$, which indeed imposes restrictions on the UAV's locations with a given BS's location.

Generally speaking, the antenna gain of the BS is not isomorphic in the three-dimensional space. The antenna of the BS is ominidirectional in horizon while the vertical antenna pattern of the BS is directional\cite{GloMMA,ICCMMA}. The vertical antenna beamwidth and down-tilt angle of the BS are given as ${\theta _B}$ and ${\theta _t}$, respectively, as shown in Fig. \ref{Channel}. It can be observed that a UAV is served by either main lobe or side lobe of the antenna. The main lobe and side lobe gains of the antenna are denoted by $g_m$ and $g_s$, respectively. Then, we have
\begin{align}
\label{CGain}
g_t = \left\{ \begin{array}{l}
{g_m},r \in \left\{ {\left. r \right|{h_{BS}} - {r}\tan \left( {{\theta _t} + {{{\theta _B}} \mathord{\left/
 {\vphantom {{{\theta _B}} 2}} \right.
 \kern-\nulldelimiterspace} 2}} \right) < {h_{UAV}} < {h_{BS}} - {r}\tan \left( {{\theta _t} - {{{\theta _B}} \mathord{\left/
 {\vphantom {{{\theta _B}} 2}} \right.
 \kern-\nulldelimiterspace} 2}} \right)} \right\}, \\
{g_s}, \text{otherwise}.
\end{array} \right.
\end{align}
Typically, the main lobe gain is much higher than the side lobe, i.e., $g_m \gg g_s$.

There are several differences between the ground-to-UAV channel and UAV-to-ground channel, one of which results from the antenna radiation patterns of BSs and ground users. The ground users generally employ omnidirectional antennas due to hardware constraints. There is no obvious gap between the spatial signals from different directions. The BSs can support complex hardware setups, and thus the antenna is ominidirectional in horizon and directional in vertical, which causes main lobe and side lobe. When the UAV is served by a BS (as an aerial user experienced the ground-to-UAV channel), it needs to distinguish the main lobe and the side lobe. In addition to the difference of antenna radiation patterns, as BSs are usually higher than terrestrial users, there are fewer obstacles between the BS and UAVs. Therefore, the probability of having LOS links in ground-to-UAV channel is greater.
\subsection{Interference Characteristics}
It can be seen from (\ref{Probability}) and (\ref{CGain}) that the received signal power of a UAV is tightly related to its location. The interference experienced by the UAV generally is more serious and complex than a ground node due to the signal transmission characteristics. In particular, the probability of having LOS interference link for the UAV is larger than a ground node since there are fewer obstacles in the sky. Because of the tremendous LOS interference, the number of BSs inducing interference to the UAV is also more. Therefore, the experienced interference will be more serious compared with ground nodes. According to the antenna characteristic of the BS, the interference can be classified into two types: Main lobe interference (e.g., UAV-2 and 3 in Fig. \ref{Channel}) and side lobe interference (e.g., UAV-1 and 4 in Fig. \ref{Channel}). It can be observed from Fig. \ref{Channel} that a UAV (i.e., UAV-3 in Fig. \ref{Channel}) served by side lobe of a nearby BS is more likely to receive the main lobe interference from the remote BSs. In practice, it is desired that the UAV is served by the main lobe while interfered by side lobe.
\section{Design and Optimization Considerations}
This section focuses on the design and optimization considerations specifically for network-connected UAV communications, including antenna design, NOMA communications, as well as network selection and association optimization.
\subsection{Antenna Design}
Known from the analysis in section III, the antenna radiation pattern of the BS is a crucial factor that influences the system performance. The existing antennas of BSs are primarily designed to serve the terrestrial users, or some users with a relatively low altitude. Consequently, the angle of the antenna is inevitably downward. However, the UAV may be higher than the BS, which results in significant antenna gain reductions for the UAV. The antenna design must take into account both the ground user and the aerial UAV, whereas it is usually difficult to serve them both efficiently and fairly.

\emph{Antenna altitude:} The received signal strength is closely related to the distance between the UAV and the BS, where the antenna altitude is a vital factor. Because it not only plays a primary role in the probability of having LOS link, but also determines the large-scale path loss. There are many factors to be considered in optimizing the antenna altitude, such as the height of the UAV. On one hand, UAVs are served better with high altitude. However, high altitude will cause great large-scale path loss to the ground users. Since the BS initially and mainly serves the terrestrial users, the antenna altitude must be designed considering the requirements of terrestrial users.

\emph{Antenna beamwidth:} The coverage of the BS, to a large extent, relies on the antenna beamwidth, i.e., the main lobe. Increasing the antenna beamwidth can enlarge the coverage, but this would reduce the antenna gain of the main lobe. Moreover, the antenna beamwidth also determines the type of interference. Although the coverage can be enhanced with larger antenna beamwidth, the UAV would experience more severe interference from main lobe.

\emph{Antenna downtile:} Antenna downtile indicates the direction of antenna propagation. Similar with the antenna beamwidth, it also has a vital impact on the coverage and interference.

Besides the antenna of the BS, the antenna of the UAV should also be carefully designed. Specifically, the directional antenna of the UAV is also a factor that can be used to improve the system performance. For example, we can reduce the beamwidth to mitigate the received interference.
\subsection{NOMA for Network-Connected UAV Communications}
Although there are many challenges produced by the antenna characteristics of the BS, some opportunities can also be founded. One of them is the application of the promising multi-user access scheme, nonorthogonal multiple access (NOMA) with successive interference cancellation (SIC)\cite{NOMA2}. Unlike traditional orthogonal multiple access (OMA) schemes that multiple users occupy orthogonal resource, such as time division multiple access (TDMA) and  orthogonal frequency division multiple access (OFDMA), multiple users in NOMA technique can be assigned to the same frequency-time resource so as to improve the spectrum efficiency. Before decoding their own information, the users with better channel conditions first employ SIC technique to remove the information intended for other users in NOMA\cite{NOMA3}. The basis of NOMA implementation relies on the difference of channel conditions among users. This difference is more remarkable in the three dimensional space for the network-connected UAV communications. For example, two nearby UAVs may be within the coverage of the main lobe and side lobe, respectively. Therefore, the channel conditions are distinctly different. Since NOMA technique enables multiple UAVs to access the same time-frequency resource block simultaneously, it advances the realization of massive access.
\subsection{Network Selection and Association Optimization}
Multi-BS coverage provides additional opportunities for network-connected UAV communications, but also with challenges. Specifically, a UAV can be served by a nearby BS and may also be in the coverage of multiple distant BSs. The UAV can simply choose the nearest BS to access network. It can also compare the received signal strength, and further select the BS that provides the strongest signal. This raises the network selection and UAV association issues that usually involve discrete variables. It is particularly difficult to be addressed considering that there are massive UAVs, where the computational complexity will increase exponentially with the number of UAVs. Moreover, in order to achieve better performance, both of them are usually jointly optimized with other issues, such as power allocation. In the network-connected UAV communications, it also involves other controllable variables, such as the location and trajectory of the UAV.
\section{Case Study}
Under the proposed network-connected UAV communications framework, many other problems still need to be investigated, such as trajectory planning and energy efficiency. In this section, we study two specific design cases for UAV association and energy efficiency optimization.
\subsection{Case Study I:  Association Methods for Network-Connected UAV Communications}
Consider a 1000 $\times$ 1000 $\text{m}^2$ region, where a UAV is served by the terrestrial BS and in the coverage of multiple BSs. The transmit power and height of the BS are -6 dBw and 30 m. The path losses at the reference distance $d_0 = 1~\text{m}$ are ${G_L} = -32.9~\text{dB}$ for the LOS link and ${G_N} = -41.1~\text{dB}$ for the NLOS link. The path loss exponents are ${{\alpha _L}} = 2.09$ and ${{\alpha _N}} = 3.75$, respectively. The environment parameters $a = 0.3$, $b = 500$, and $c = 15$\cite{GloMMA,ICCMMA}. Without other explanations, the beamwidth and downtitlt angle of the BS's antenna are ${\theta _B} = {30^o}$ and ${\theta _t} = {8^o}$, respectively. The gains of main lobe and side lobe are $g_m=10$ and $g_s=0.5$\cite{GloMMA,ICCMMA}. For the association between the UAV and the BS, two methods are considered: Closest association and strongest association. In particular, the UAV is associated with the BS to which it is closest in closest association and from which it receives the strongest signal in strongest association.
\begin{figure}
\centering
\subfigure[The SINR versus the height of the UAV]{
\includegraphics[width=0.4\textwidth]{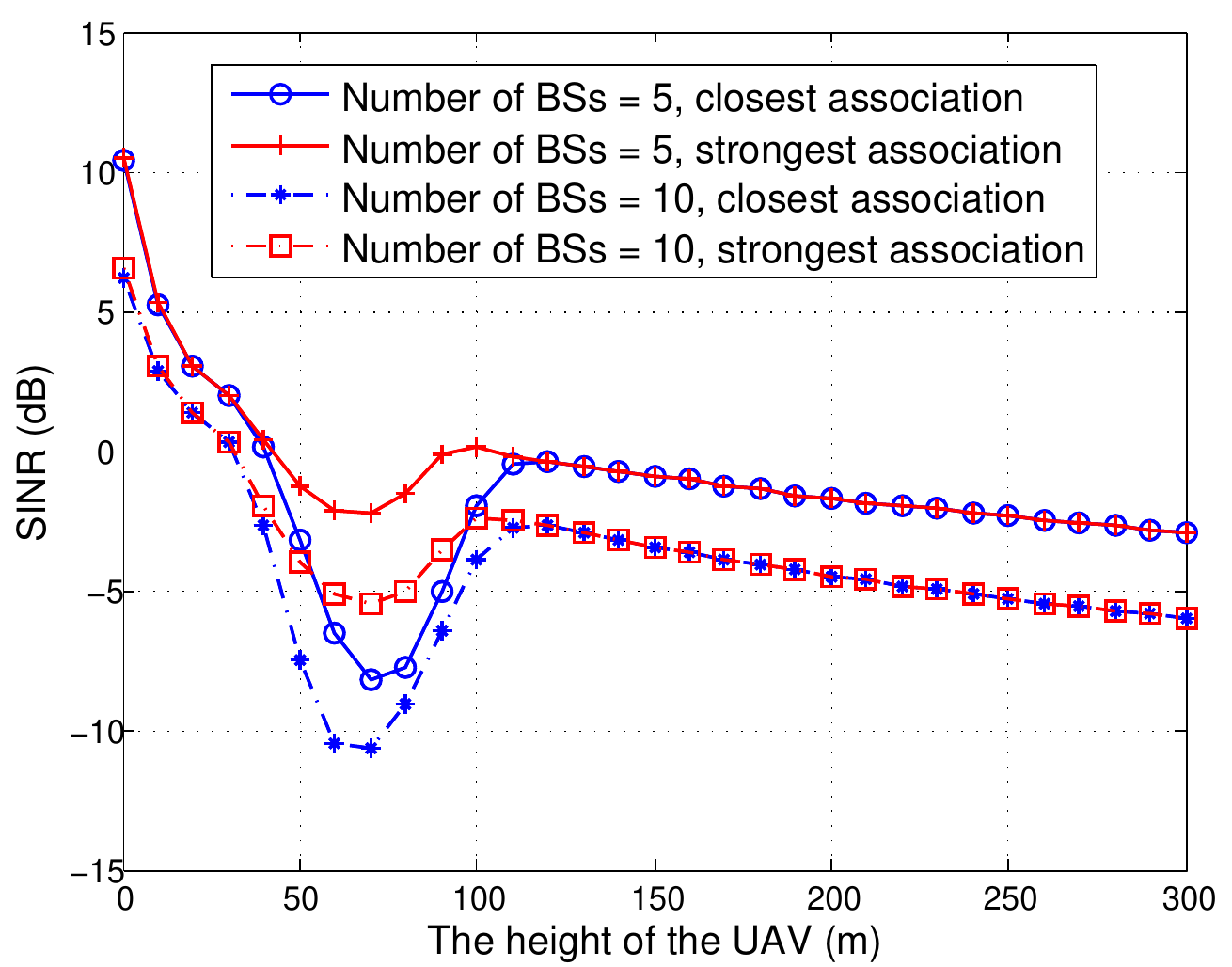}}
\subfigure[The SINR versus the MSR]{
\includegraphics[width=0.4\textwidth]{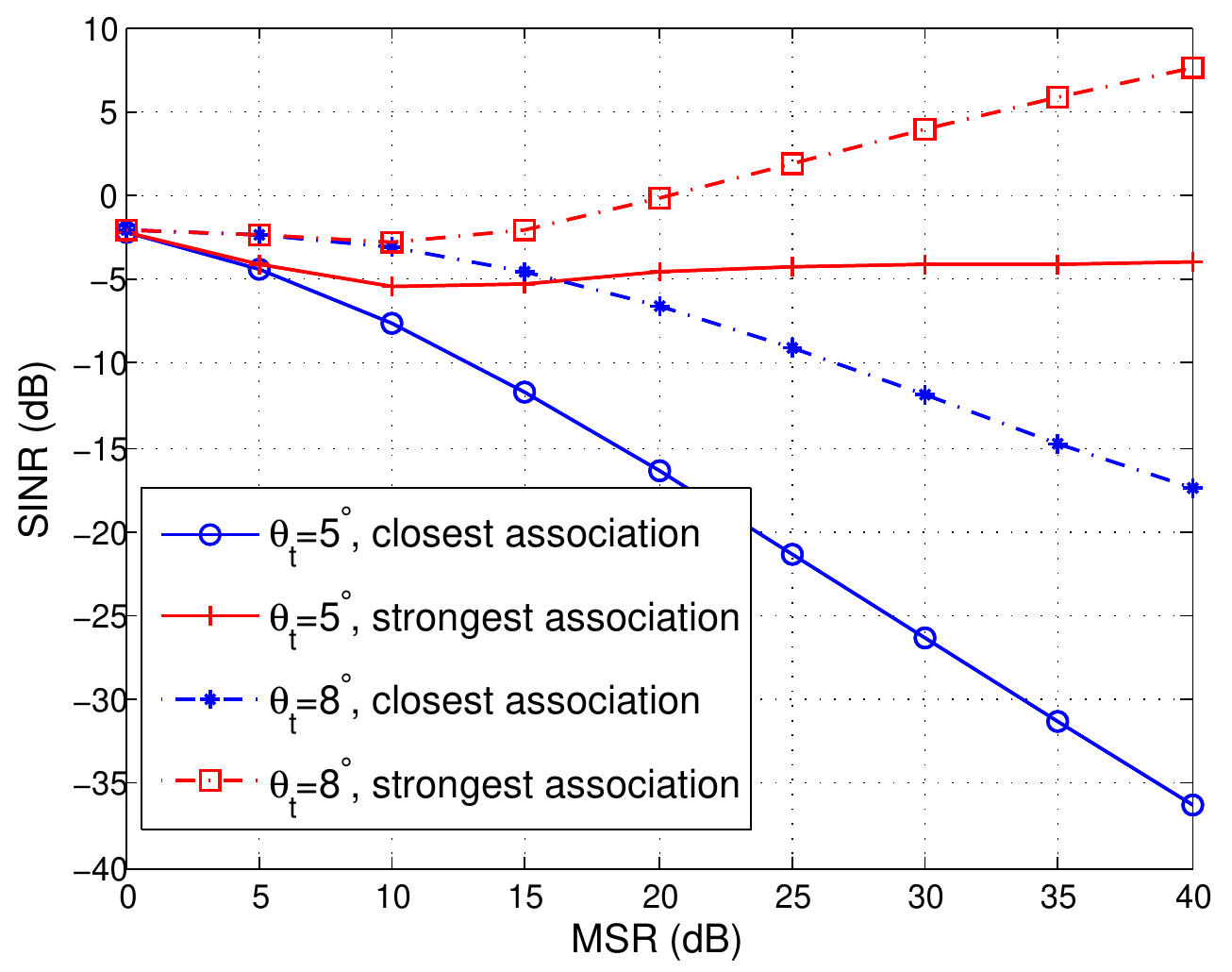}}
\caption{The SINR versus the height of the UAV and the MSR under different association methods.}
\label{CaseI}
\end{figure}

In Fig. \ref{CaseI}(a), the signal to interference plus noise ratio (SINR) of the UAV under different association methods is plotted versus the height of the UAV. We can find that it is not monotonous for the variation tendency between the SINR and the height of the UAV. The whole process can be divided into three phases. The SINR first decreases as the height increases. This is because the UAV suffers severe main lobe interference introduced by other BSs. Moreover, the number of BSs producing LOS interference to the UAV is also more. Then, as the height continues to increase, the main lobe interference becomes the side lobe interference, which improves the SINR. In the third phase, since the UAV is far away from the BS that serves it, the signal received by the UAV is weakened. Consequently, the SINR decreases in this case. In summary, the height of the UAV has an significant impact on the system performance, which needs to be carefully investigated. It can be also observed that the achieved results by two association methods are not always the same, which means that the BS providing the strongest signal is not always the one that is closest to the UAV. This is due to the reason that the signal received from the main lobe of a distant BS may be stronger than the signal received from the side lobe of a nearby BS.

Fig. \ref{CaseI}(b) illustrates the SINR versus the main lobe to side lobe ratio (MSR), where the number of BSs is 10 and the height of the UAV $h_{BS}= 100~\text{m}$. It is first observed that the SINR in closest association decreases as the MSR increases. This is due to the fact that the signal power received from the side lobe in closest association becomes smaller compared with the interference from the main lobe. Additionally, small downtile angle ${\theta _t}$ results in low SINR with a given beamwidth ${\theta _B}$. The reason is that the antennas of the BSs tend to point to the aerial UAV in the case of small downtile angle. Therefore, it causes more serious interference to the UAV. These results imply the necessity of implementing antenna optimization in order to realize the potentials of network-connected UAV communications.
\subsection{Case Study II:  Energy Efficiency for Network-Connected UAV Communications}
Generally speaking, the signal received by the celledge user is weak, however, with serious interference, which results in low SINR. In this context, a UAV can serve the celledge user to enhance the terrestrial communications\cite{CellularZY2}. We consider a UAV flies circularly above an area of 1000 m in radius, where ground nodes are equally distributed and communicate with the UAV in a cyclical time-division manner\cite{wclLJ}. The UAV energy consumption with steady circular flight is given by\cite{TWCZY}
\begin{align}
E\left( V \right) = T\left[ {\left( {{c_1} + \frac{{{c_2}}}{{{g^2}{r^2}}}} \right){V^3} + \frac{{{c_2}}}{V}} \right],
\end{align}
where $g = 9.8~\text{m}/\text{s}^2$ is the gravitational acceleration, $c_1=9.26\ast10^{-4}$ and $c_2=2250$ are parameters related to the UAV's weight, wing area, air density, etc., $r$ is the flight radius, $T$ is the flight time, and $V$ is the UAV's speed. The height and transmit power of the UAV are 100 m and 30 dBm.

\begin{figure}
\centering
\subfigure[Delay]{
\includegraphics[width=0.4\textwidth]{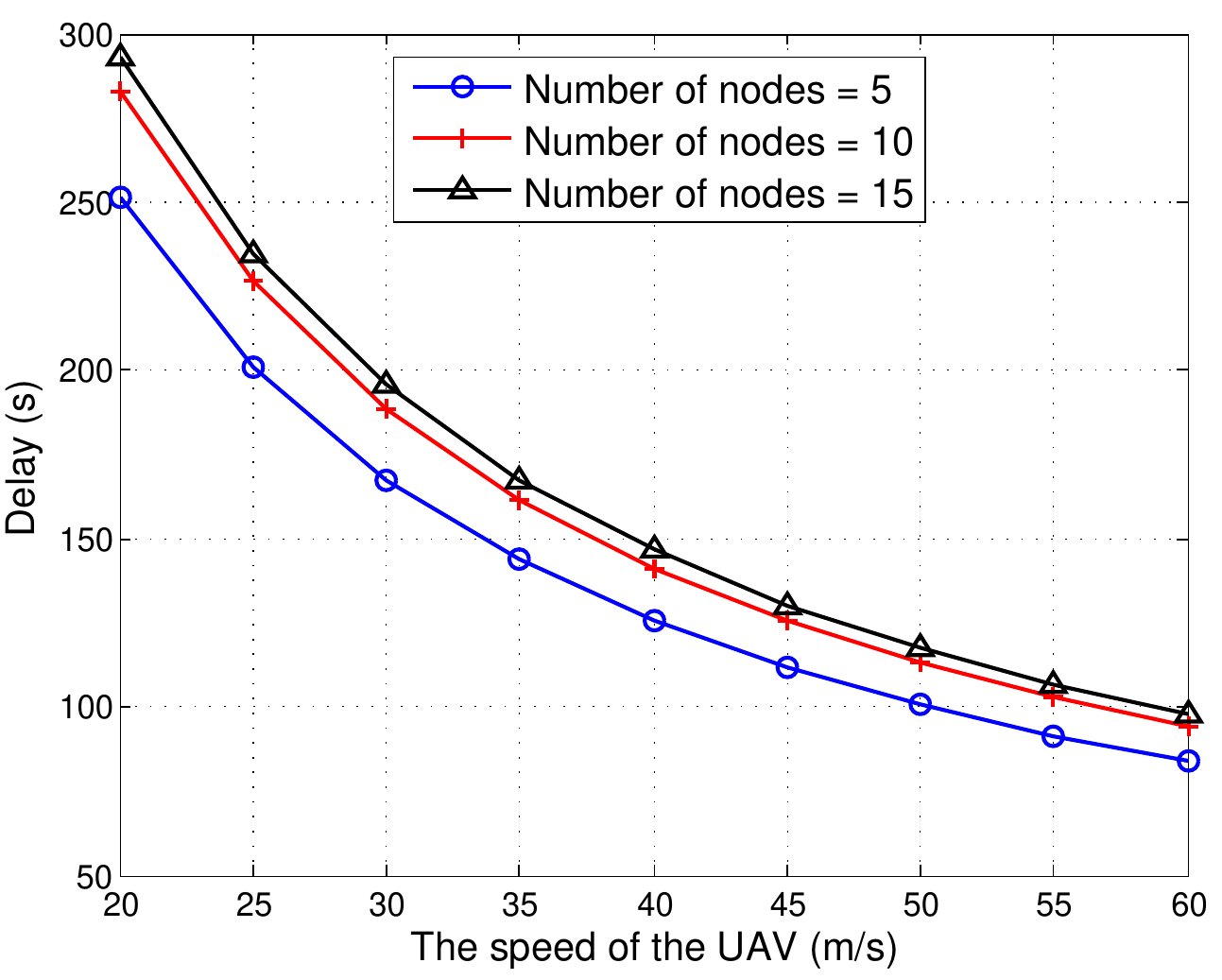}}
\subfigure[Energy efficiency]{
\includegraphics[width=0.4\textwidth]{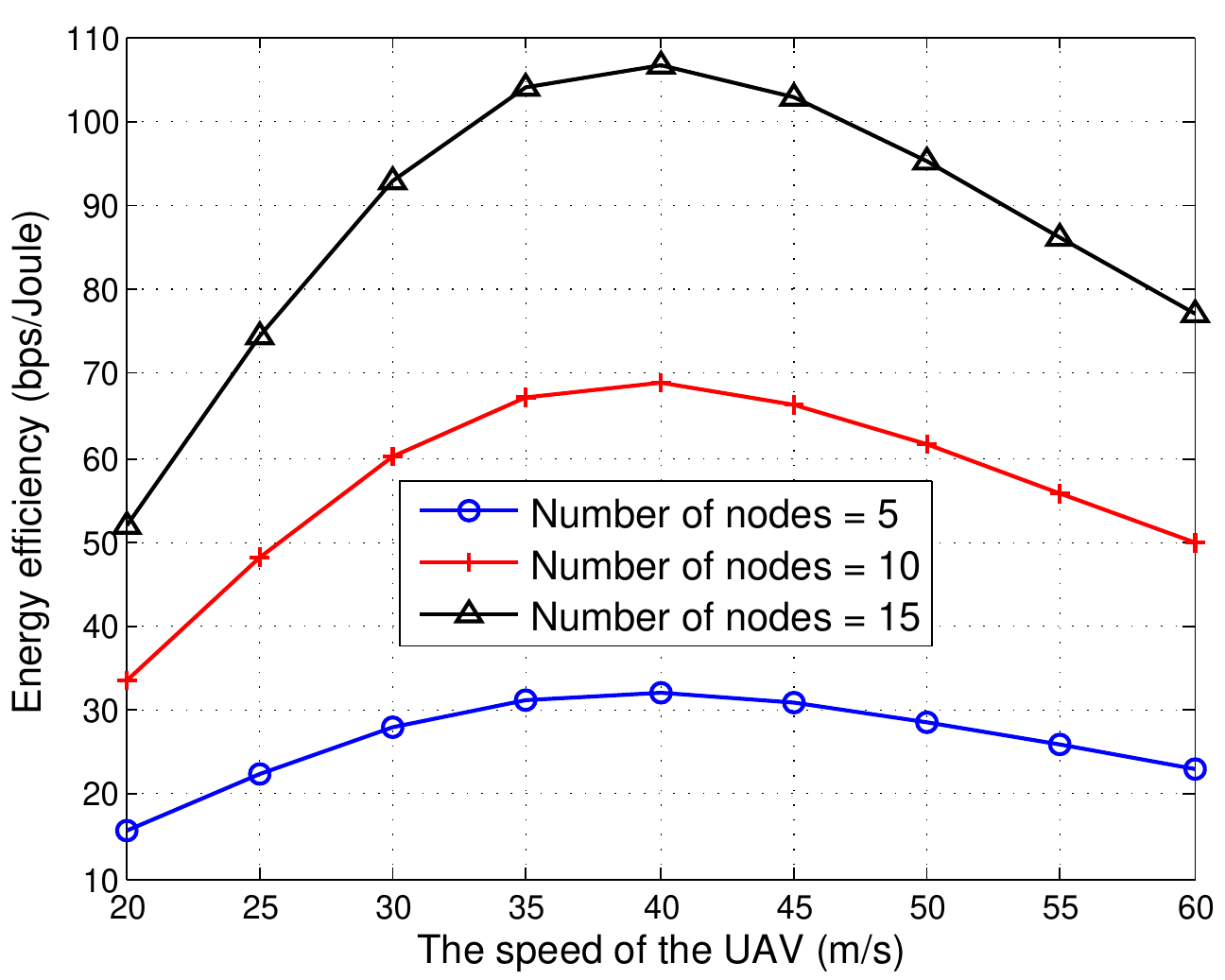}}
\caption{The delay and energy efficiency versus the speed of the UAV.}
\label{fig1}
\end{figure}
Fig. \ref{fig1}(a) illustrates the delay versus the speed of the UAV. The delay is defined as the longest time that the node is not served. It can be seen that higher speed and/or less number of ground nodes is benefical for reducing the delay. The energy efficiency versus the speed of the UAV is shown in \ref{fig1}(b), where the noise power is -174 dBm/Hz and the bandwidth is 1 MHz. The energy efficiency is defined as the achievable throughput to consumed energy ratio. Notice that the communication-related energy is ignored since it is much smaller than that used to support the UAV's mobility\cite{TWCZY}. It can be observed that, unlike delay, the energy efficiency may decrease with an increasing speed. The reason is that the consumed energy dramatically increases with high speed.
\section{Conclusions}
In this article, we studied the use of network-connected UAV communications as a compelling solution to achieve high-capacity information transmission and ultra-reliable UAV remote command and control. The aim was to elaborate the design aspects and open issues in network-connected UAV communications. In particular, we first discussed the use cases of UAVs and the associated requirements. Then, we proposed a flexible architecture for network-connected UAV communications. Subsequently, the signal transmission and interference characteristics were theoretically analyzed. Further, we investigated the design and optimization considerations, including antenna design, NOMA communications, as well as network selection and association optimization. Finally, case studies were provided to show the feasibility of network-connected UAV communications. We firmly believe this important area will be a fruitful research direction, and we have just touched the tip of the iceberg. We hope this article will stimulate much more research interest.

%

\ifCLASSOPTIONcaptionsoff
  \newpage
\fi

\end{document}